\documentclass[10pt,preprint]{aastex}
\usepackage{graphicx}
\usepackage{subfigure}

\begin{document}

\title{GAMER with out-of-core computation}

\author{Hsi-Yu Schive\altaffilmark{1,2,3}, Yu-Chih Tsai\altaffilmark{1}, \& Tzihong Chiueh\altaffilmark{1,2,3}}

\altaffiltext{1}{Department of Physics, National Taiwan University, 106, Taipei, Taiwan, R.O.C.; Email: b88202011@ntu.edu.tw (Hsi-Yu Schive)}
\altaffiltext{2}{Center for Theoretical Sciences, National Taiwan University, 106, Taipei, Taiwan, R.O.C.}
\altaffiltext{3}{Leung Center for Cosmology and Particle Astrophysics, National Taiwan University, 106, Taipei, Taiwan, R.O.C.}

\maketitle

\begin{abstract}
\emph{GAMER} is a GPU-accelerated Adaptive-MEsh-Refinement code for astrophysical simulations. In this work, two further extensions of the code are reported. First, we have implemented the MUSCL-Hancock method with the Roe's Riemann solver for the hydrodynamic evolution, by which the accuracy, overall performance and the GPU versus CPU speed-up factor are improved. Second, we have implemented the out-of-core computation, which utilizes the large storage space of multiple hard disks as the additional run-time virtual memory and permits an extremely large problem to be solved in a relatively small-size GPU cluster. The communication overhead associated with the data transfer between the parallel hard disks and the main memory is carefully reduced by overlapping it with the CPU/GPU computations.

\keywords{gravitation, hydrodynamics, methods: numerical}
\end{abstract}

\section{Introduction}

Novel use of the graphic processing unit (GPU) has becoming a promising technique in the computational astrophysics. The applications that have been reported include purely hydrodynamics, magnetohydrodynamics, gravitational lensing, radiation transfer, direct N-body calculation, particle-mesh method, hierarchical tree algorithm, and etc. Typically, one to two order of magnitudes performance improvements were reported \citep[e.g.,][]{Schive2008}.

\citet{Schive2010} presented the first multi-GPU-accelerated Adaptive-MEsh-Refinement (AMR) code, named \emph{GAMER}, which is dedicated for high-resolution astrophysical simulations. A GPU hydrodynamic solver and a GPU Poisson solver have been implemented in the code, while the AMR data structure is still manipulated by CPUs. An overall performance speed-up up to 12x was reported. In this work, two further extensions are implemented into \emph{GAMER}, namely, the MUSCL-Hancock method \citep{Toro2009} with the Roe's Riemann solver \citep{Roe1981} for the hydrodynamic evolution, and the out-of-core computation. The latter uses parallel hard disks to increase the total amount of available virtual memory. By integrating the high computation performance of GPUs and the out-of-core technique, it provides an extremely efficient solution to increase both the simulation problem size and performance of the AMR simulations.

\section{Extension I: hydrodynamic solver}

\begin{figure}[h]
     \centering
    \scalebox{1.0}{%
    \includegraphics[width=7cm]{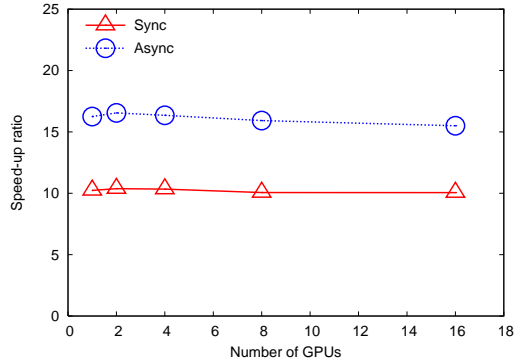}
    }
    \caption{Overall performance speed-up versus the number of GPUs using the relaxing TVD scheme. The circles and triangles show the results with and without the concurrent execution between CPU and GPU, respectively.}
    \label{Fig_SpeedupRatio_Cosmology_RTVD}
\end{figure}

In the previous work \citep{Schive2010}, the second-order relaxing total variation diminishing (TVD) method \citep{TP2003} has been adopted, in which the three-dimensional evolution is achieved by the dimensional splitting method. Figure \ref{Fig_SpeedupRatio_Cosmology_RTVD} shows the performance speed-up versus the number of GPUs, in which we compare the performance using the same number of GPUs (Tesla T10 GPU) and CPU cores (Xeon E5520). The simulations are conducted in the GPU cluster installed in the National Astronomical Observatories, Chinese Academy of Sciences. We also compare the results with and without the concurrent execution between CPU and GPU, and a maximum speed-up up to 16.5x is demonstrated when the concurrency is enabled. We also notice that the speed-up factor only decreases slightly to 15.5x in the 16 GPUs/CPUs test, indicating that the network time is nearly negligible. Timing measurements show that the MPI data transfer takes less than 2\% of the total simulation time.

To further enhance the capability of the code, in this work we have implemented a new GPU hydrodynamic solver based on the MUSCL-Hancock method. This approach contains four steps, namely, the spatial data reconstruction, the half-step prediction, the Riemann solver, and the full-step update. We apply the unsplit finite volume method for the three-dimensional evolution, and the Roe's solver is adopted for the Riemann problem. Comparing with the second-order relaxing TVD scheme, the MUSCL-Hancock method has two main advantages for the AMR+GPU implementation. First, it requires only a five-point stencil in each spatial direction, while the relaxing TVD scheme requires seven points. Since the operation of preparing the ghost-zone data is conducted by CPU, which has been shown to be more time expensive than the GPU hydrodynamic solver \citep{Schive2010}, reducing the size of stencil can directly lead to significant improvement of the overall performance. Second, the MUSCL-Hancock method has higher arithmetic intensity, and hence is more GPU-friendly. Factor of 55x performance speed-up is measured by comparing the GPU and CPU versions of this method.

\begin{figure}[h]
    \centering
    \scalebox{1.0}{%
    \includegraphics[width=7cm]{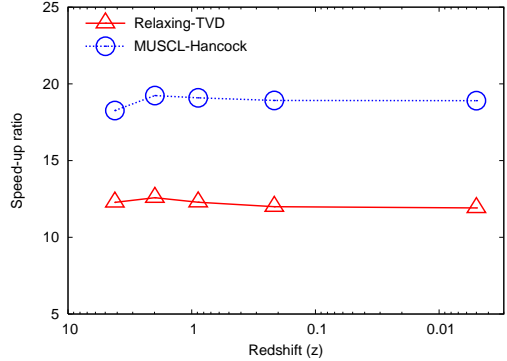}
    }
    \caption{Overall performance speed-up in purely baryonic cosmological simulations using the MUSCL-Hancock scheme (circles) and the relaxing-TVD scheme (triangles), respectively.}
    \label{Fig_RTVD_vs_MUSCL-H}
\end{figure}

Figure \ref{Fig_RTVD_vs_MUSCL-H} compares the overall performance speed-up in purely baryonic cosmological simulations using the two different hydrodynamic schemes. The same gravity solver is adopted in the two cases. The performance is measured by using one NVIDIA GeForce 8800 GTX GPU and one AMD Athlon 64 X2 3800 CPU core. Clearly, the MUSCL-Hancock scheme achieves a superior performance improvement, in which a speed-up of 19.2x is demonstrated. Moreover, although in the CPU-only runs the MUSCL-Hancock method is more time-consuming, it is not the case when the GPU-acceleration is activated. Timing experiments show that the total execution time is actually reduced when we replace the GPU relaxing-TVD solver by the GPU MUSCL-Hancock solver, which results from the less computing time required for the ghost-zone preparation in CPU and the more efficient GPU kernel.

\section{Extension II: out-of-core computation}

In \emph{GAMER}, we have demonstrated that the performance of the AMR simulations can be highly improved by using GPUs. However, the simulation size is still limited by the total amount of main memory. To alleviate this limitation, we further implement the out-of-core technique, by which only a small portion of the simulation data need to be loaded into the main memory while the rest of data remain stored in the hard disks. To increase the total I/O bandwidth in a single node, we evenly distribute the data in eight hard disks and perform the data transfer between the main memory and the eight disks concurrently. By doing so, a maximum bandwidth of 750 MB/s is achieved.

\begin{figure}[h]
    \centering
    \includegraphics[width=3.7cm]{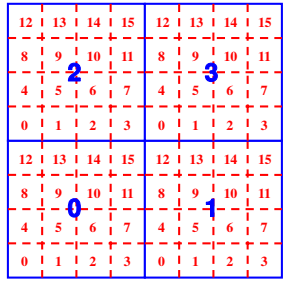}
    \caption{Domain decomposition of the parallelized out-of-core computation. The solid blue lines represent the sub-domain boundaries between different MPI ranks in different computing nodes, and the dashed red lines represent the sub-domain boundaries between different out-of-core ranks in the same computing node. The blue and red numbers stand for the MPI ranks and the out-of-core ranks, respectively.}
    \label{Fig_OOC_DomainDecomposition}
\end{figure}

The parallelization in \emph{GAMER} is based on the rectangular domain decomposition. To perform the out-of-core computation in a multi-node system, we let each computing node to work on a group of nearby sub-domains, and each of which will be assigned a different out-of-core rank that is similar to the concept of the MPI rank. Figure \ref{Fig_OOC_DomainDecomposition} shows a two-dimensional example of the domain decomposition. Different sub-domains within the same node are always evaluated sequentially, while sub-domains in different nodes can be evaluated in parallel. In each node, only the data of the sub-domain being advanced are loaded from the hard disks to the main memory. After the targeted sub-domain is advanced by one time-step, the updated data will be stored back to the hard disks, and the data of the next targeted sub-domain will be loaded into the main memory. Furthermore, to improve the efficiency of the out-of-core computation, the hard disk I/O time for one out-of-core rank is arranged to be overlapped with the computation for a different out-of-core rank in the same node.

Updating the buffer data of each sub-domain requires transferring data in between adjacent sub-domains. However, since different out-of-core ranks within the same computing node are calculated sequentially, we need a data transfer mechanism different from the MPI implementation. To this end, we have implemented two functions named \emph{OOC\underline{ }Send} and \emph{OOC\underline{ }Recv}, which are similar to the MPI functions \emph{MPI\underline{ }Send} and \emph{MPI\underline{ }Reve}, but use the hard disks as the data exchange buffer. For example, to send data from the out-of-core rank A to rank B, the former first invokes the function OOC\underline{ }Send to store the transferring data in the hard disks. Afterward, rank B can invoke the function OOC\underline{ }Recv with the correct data tag to load the transferring data from the hard disks, thus completing a single data transferring operation. On the other hand, the data transfer between different computing nodes is still accomplished by using the MPI functions.

\begin{figure}[h]
    \centering
    \subfigure[CPU]{
        \includegraphics[width=6cm]{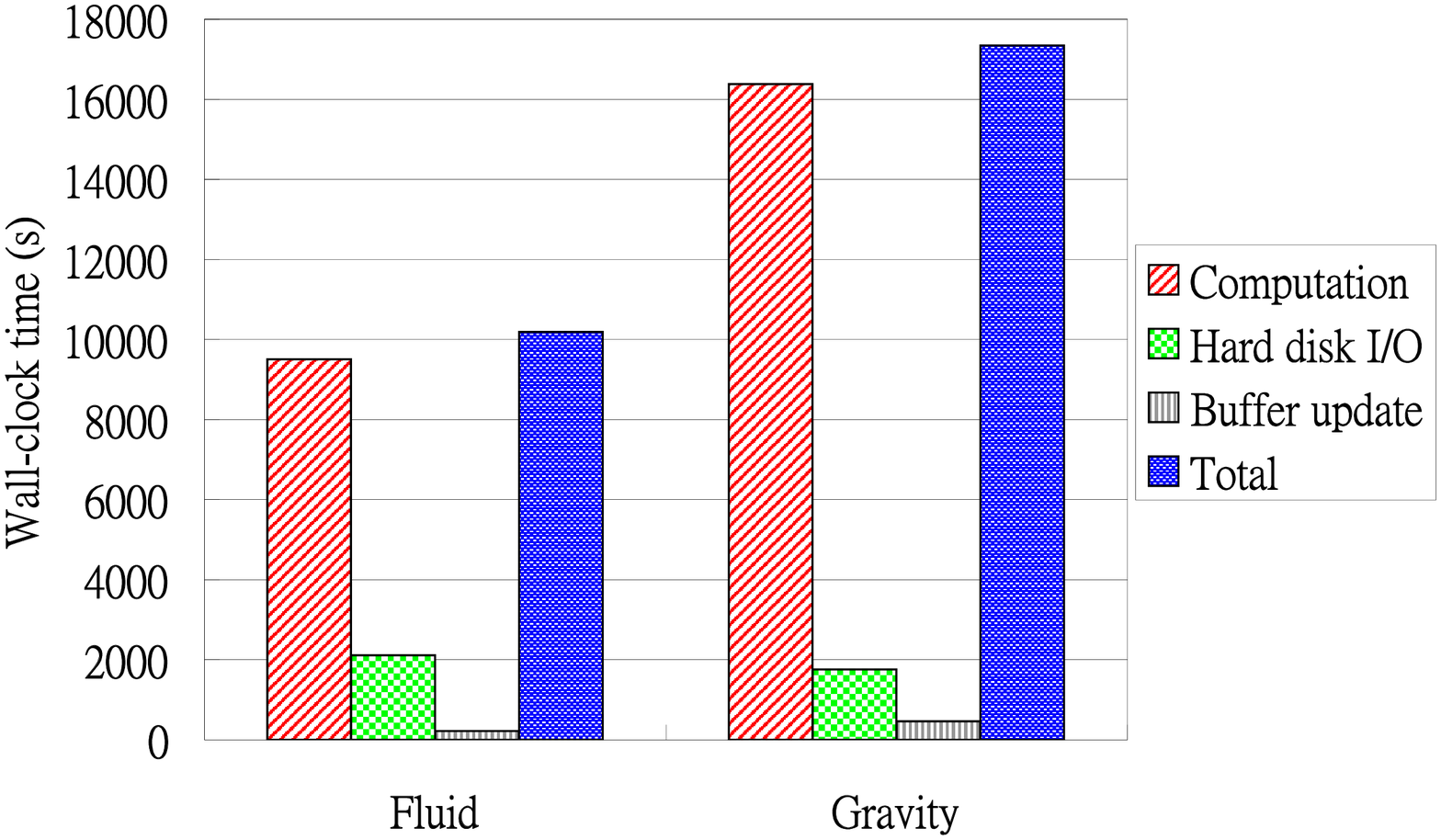}
        \label{Fig_OOC_Performance_AMR512_CPU}
    }
    \hspace{0.1cm}
    \subfigure[GPU]{
        \includegraphics[width=6cm]{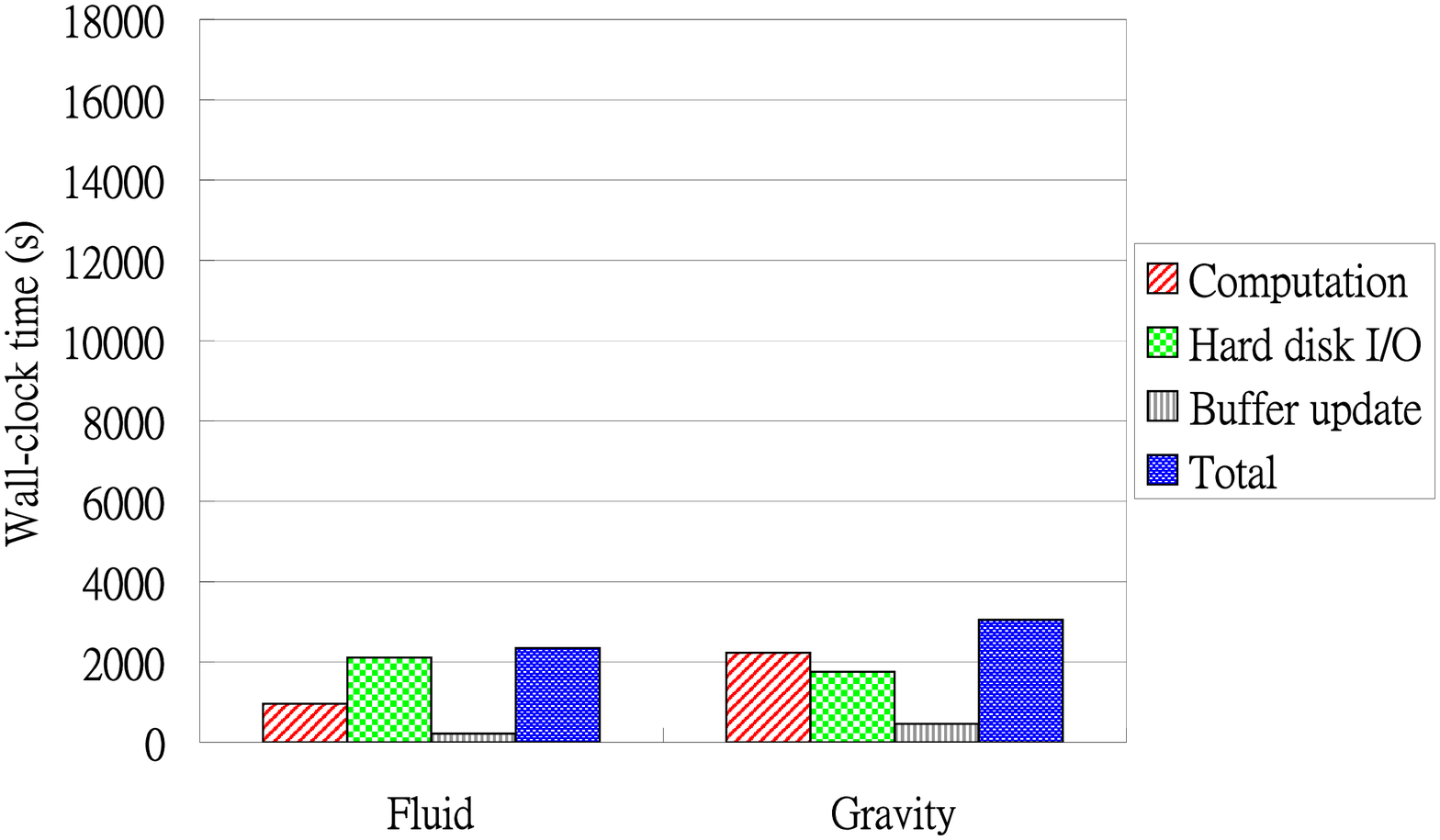}
        \label{Fig_OOC_Performance_AMR512_GPU}
    }
    \caption{Performance of the fluid and gravity solvers in the out-of-core AMR simulations. The right and left panels show the results with and without the GPU-acceleration, respectively.}
    \label{Fig_OOC_Performance_AMR512}
\end{figure}

To test the performance, we conduct single-node simulations with the $512^3$ root level and five refinement levels, giving $16,384^3$ effective resolution. The total memory requirement is about 100 GB. By dividing the simulation domain into 64 sub-domains, the amount of memory actually allocated is only 3 GB. The performance is measured by using one NVIDIA GeForce 8800 GTX GPU and one Intel i7-920 CPU core. Figure \ref{Fig_OOC_Performance_AMR512} shows the timing measurements of the hydrodynamic and gravity solvers with and without the GPU acceleration. In the CPU-only case, the data I/O time is always much shorter than the computation time. In the case with GPU acceleration, the performance of the hydrodynamic solver is dominated by the data I/O, while that of the gravity solver is still dominated by the CPU/GPU computation. Also note that in each case, the total elapsed time is significantly shorter than the sum of the computation time and the I/O time, indicative of efficient overlap between computation and data I/O. We conclude that the out-of-core computation is reviving and can potentially be a powerful vehicle to deliver the optimal performance of a GPU cluster.

\end{document}